\documentclass[twocolumn,prd]{revtex4}

\usepackage{amssymb}

\input{epsf}

\begin{document}

\title{\bf Stability of pentaquarks with a two- plus three-body chromoelectric interaction}

\date{\today}
\author{Fl. Stancu\footnote{E-mail address: fstancu@ulg.ac.be}}
\affiliation{University of Li\`ege, Institute of Physics B5, Sart Tilman,
B-4000 Li\`ege 1, Belgium}

\begin{abstract}

We study the stability of pentaquarks within a schematic model based on SU(3) color symmetry,
by taking into account algebraic arguments leading to a   
chromoelectric interaction containing two- and three-body parts. It has already been 
proven that such an interaction  can influence the spectrum of ordinary baryons
and the stability of tetraquarks and hexaquarks. Here we discuss its role in the stability 
of pentaquarks against strong decays.
\end{abstract}

\maketitle

\section{Introduction}

First mentioned by Gell-Mann in his
original paper on the quark model of baryons and mesons
\cite{GellMann:1964nj},
the existence of exotic hadrons as systems with more than three quarks
(antiquarks) has become an important 
challenge for QCD-inspired quark models. 
In practice the problem has been initiated by Jaffe who noticed that the color-magnetic
hyperfine interaction  produced a stable   
$uuddss$ hexaquark, called the $H$-particle, \cite{Jaffe:1976yi}
and stable $q^2 {\overline q}^2$ systems \cite{Jaffe:1976ig}

Ten years later,
still based on the one-gluon exchange model Gignoux, Silvestre-Brac and 
Richard \cite{Gignoux:1987cn} and independently Lipkin \cite{Lipkin:1987sk}
found that the strange anticharm pentaquarks $uuds {\overline c}$ and $udds {\overline c}$
are also stable against strong decays. To be stable, these particles
had to contain a heavy flavor. Generally,  one
expects an increase in the stability if a multiquark system contains
heavy flavors $Q = c$ or $b$. The above pentaquarks have negative parity,
i. e. the parity of the antiquark.  
Later on,  it was shown \cite{Stancu:1998sm} that the 
Goldstone-boson exchange model favors positive parity pentaquarks 
and stabilizes systems of type 
$uudd {\overline Q}$, without strange quarks. 
On the experimental side the H1 Collaboration has found evidence for a narrow 
resonance interpreted as a $u u d d \bar c$ pentaquark \cite{Aktas:2004qf},
so far unconfirmed by other experiments.
An earlier experiment of the E791 Collaboration \cite{Aitala:1999ij}
failed to obtain statistically significant signals for a narrow strange 
anticharm pentaquark. An early short review on the
stability of multiquark systems can be found, for example, 
in Ref. \cite{Stancu:1999xc}.
The physics of exotic hadrons has recently been reviewed in the spirit of 
quark models \cite{Richard:2016eis} in a more advanced picture with
an expansion in the Fock space to allow additional quark-antiquark pairs 
in the wave function of baryons.
 
The recent observation of the hidden-charm pentaquarks $P_c(4380)$ with $J^P = 3/2^-$
and $P_c(4450)$ with $J^P = 5/2^+$ by the LHCb collaboration \cite{Aaij:2015tga}
has triggered a new strong wave of interest in pentaquarks.
Competing models have already been proposed, see Ref. \cite{Burns:2015dwa} for a
recent review on the phenomenology of pentaquarks. 

The stability of a few quark systems, 
like the stability of  a few-charge systems depends on the masses of the fermions which are involved \cite{Richard:2017vry}.
Most of the results discussing  the mass dependence are related to heavy-light tetraquarks of type
$ Q Q \bar q \bar q$ as a function of the quark mass ratio $M/m$ \cite{Zouzou:1986qh,Brink:1998as,Janc:2004qn}.
To reach stability one needs a rather large $M/m$. 
The results are model dependent and are also influenced by the performance of the variational method,
the most elaborate being that of Ref. \cite{Janc:2004qn}, where the stability of the $ c c \bar u \bar d$
was achieved. The study of the $ Q \bar Q q \bar q$ is more complicated because the lowest
threshold is $ Q \bar Q + q \bar q$ [13].
In constituent quark models the chromomagnetic interaction also plays an important role.
But in the limit of very heavy quarks this interaction vanishes and the result depends on a pure 
chromoelectric interaction. The kinetic energy decreases by increasing the constituent mass,
so that one gets more binding.

Here we discuss a mechanism which could influence the stability of
pentaquarks against strong decays in quark models due to a pure chromoelectric interaction, which contains a two- plus 
a three-body confining potential as suggested in Ref.  \cite{Dmitrasinovic:2001nu}. 
Usually constituent quark models contain a two-body $F_i \cdot F_j$
confining interaction only.  
Based on the algebraic argument
that SU$_C$(3) is an exact local symmetry of QCD which implies that QCD-inspired
Hamiltonian models are invariant under a global SU$_C$(3) symmetry, 
in Ref. \cite{Dmitrasinovic:2001nu}
it was suggested that the quark model Hamiltonian should be expressed 
in terms of every invariant operator of SU(3). 
This implies that the Hamiltonian should contain both two- and three-body
confining forces.
In other words, 
the two-body confining force $F_i \cdot F_j$
can be expressed in terms of the quadratic
(Casimir) operator and the three-body force (see next section)
in terms of the cubic
invariant. If added to the Hamiltonian,
the three-body force has implications on the spectrum of ordinary baryons
\cite{Dmitrasinovic:2001nu,Papp:2002xh} and on the stability of tetraquarks 
\cite{Dmitrasinovic:2001nu,Pepin:2001is,Dmitrasinovic:2003cb} and 
six-quark systems (the nucleon-nucleon system) \cite{Pepin:2001is}.
Here we discuss its effect on the stability of pentaquarks
containing both light and/or heavy quarks, inasmuch as the confining interaction 
is flavor independent. 

In Refs. \cite{Dmitrasinovic:2001nu,Papp:2002xh,Pepin:2001is} it was shown that the three-body confining
interaction 
introduced in the next section can modify the
gap between physical color singlet states and the unphysical
color octet states which plague the spectrum of a
quark model Hamiltonian containing two-body color confinement only.
For negative values of the relative strength  $c$ between the three- and two-body 
confinement interaction the color states are shifted 
high up in the spectrum. 
Then they can be neglected in calculations and
the baryon can be described as a $q^3$ state. 
For $c \geq 0 $ the color states are located in the observed region of the 
spectrum. 
Hence they have to be considered as giving rise to $(qqq)(q {\overline q})$
configurations to be admixed with $q^3$ configurations.
It has been suggested that most of the low lying baryon resonances
could have substantial $q^4 {\overline q}$ admixtures, see, for example, Ref. \cite{Helminen:2000jb}. 
The three-body force presented here can lower the energy of a 
$q^4 {\overline q}$ state close to that of a  $q^3$ state. Then one 
can expect significant admixtures of $q^4 {\overline q}$
with  $q^3$ configurations in the baryon wave functions, as mentioned in Ref.  \cite{Richard:2016eis}.

In Ref. \cite{Pepin:2001is} it was shown that the three-body confining force increases 
the coupling between physical states and hidden-color (octet-octet) states
in $q^6$ systems
if $c$ is negative and has therefore some influence on the nucleon-nucleon
interaction derived in a microscopic approach based on the quark structure
of baryons.

In the present approach the arguments are purely algebraic but a three-body confining interaction 
could possibly mimic the little information we have from QCD lattice results
\cite{Bali:2000gf,Takahashi:2002bw,Alexandrou:2001ip,Suganuma:2011ci,Bicudo:2015ofa}.
Also, it would be useful to establish some connection between flux-tube models for 
pentaquarks and the present work. 

The paper is organized as follows. In the next section we briefly
recall the three-body confining interaction of Refs. \cite{Dmitrasinovic:2001nu,Papp:2002xh,Pepin:2001is}. 
In Sec. III we introduce the 
basis color states specific to the $q^4 {\overline q}$ system
and derive a useful unitary
transformation between states obtained in two relevant coupling schemes.
In Sec. IV we present results regarding the role of the three-body
confinement interaction on the stability of pentaquarks and implications
on the spectra of ordinary baryons. 
An estimate of the effect of this interaction is made in the case of a simple model  
of  harmonic oscillator confinement.
The last section is devoted to conclusions. Useful group theoretical details
are described in Appendices \ref{fbar} - \ref{alternative}.

\section{Two- and three-body confining interactions}
Let us consider the Hamiltonian
\begin{equation}\label{HAMILT}
H = \sum\limits_{i} \frac{{\vec{p}_i}^2}{2 m_i} - \frac{\vec{P}^2}{2M} 
  + V_{2b} + V_{3b}
\end{equation}
where $\vec{p}_i$ and $m_i$ are the momentum and the mass of the quark 
(antiquark) $i$,
$\vec{P}$ and $M$ are the total momentum and the mass of the
 $q^4 {\overline q}$ system
and $V_{2b}+ V_{3b}$ is the confinement interaction.
For the 2-body confinement interaction we choose 
the form
\cite{Dmitrasinovic:2001nu}
\begin{equation}\label{V2b}
V_{2b} = \sum\limits_{i<j} V_{ij}~( \frac{7}{3} + F^{a}_{i} \cdot  F^{a}_{j})
\end{equation}
where
\begin{equation}
F^{a}_{i} = \frac{1}{2} \lambda^{a}_{i}, \hspace{0.5cm} (a = 1,...,8)
\end{equation}
is the color charge operator of the quark $i$. 
In constituent quark models only a two-body color confinement 
interaction is usually considered. The common form of its color part is $-3/2 ~F^{a}_{i} \cdot F^{a}_{j}$, which
differs from the expression above by an additional constant. As pointed out
in Ref. \cite{Dmitrasinovic:2001nu} the constant 7/3 
ensures the stability of $q {\overline q}$ pairs. One can easily 
pass from one form to the other.

The  three-body color confinement interaction proposed in Ref. \cite{Dmitrasinovic:2001nu}
has the following form 
\begin{equation}\label{3BTOTAL}
V_{3b}=V_{ijk} = {\mathcal V}_{ijk} {\mathcal C}_{ijk}
\end{equation}
where ${\mathcal V}_{ijk}$ is the radial part 
and ${\mathcal C}_{ijk}$ is the
three-body color operator. For a $q^3$ system or a multiquark $q^3$ subsystem this
has the form 
\begin{equation}\label{3Q}
{\mathcal C}_{ijk} = d^{abc}~F^{a}_{i} ~F^{b}_{j}~ F^{c}_{k},
\end{equation}
where 
$d^{abc}$ are some real constants,
symmetric under any permutation of indices \cite{Stancu:1991rc}.
The three-body color operator acting in a $q^2 {\overline q}$ subsystem
is defined as 
\begin{equation}\label{2QQBAR}
{ {\mathcal C}}_{ij {\overline k}} = -d^{abc} ~F^{a}_i
~F^{b}_j ~ {\overline F}^{c}_k
\end{equation}
where 
\begin{equation}\label{Fantiquark}
{\overline F}^a_i = -\frac{1}{2} \lambda^{a*}_i, \hspace{0.5cm} (a = 1,...,8)
\end{equation}
is the color charge operator of an antiquark.
These operators can be expressed in terms of the quadratic invariant $C^{(2)}$
and the cubic invariant  $C^{(3)}$ of SU(3) as \cite{Dmitrasinovic:2001nu,Pepin:2001is}
\begin{equation}\label{QQQ}
{\mathcal C}_{ijk} =
\frac{1}{6}~ [~  C^{(3)}_{i+j+k} - \frac{5}{2} C^{(2)}_{i+j+k} +
\frac{20}{3}~ ]
\end{equation}
and
\begin{equation}\label{QQQBAR}
{{\mathcal C}}_{ij {\overline k}} =
-\frac{1}{6} [ C^{(3)}_{i+j+ \overline k} - \frac{5}{2} C^{(2)}_{i+j}
+ \frac{50}{9} ]
\end{equation}
We recall that for a given irrep of SU(3) labelled by $(\lambda \mu)$,
the eigenvalues of these invariants are
\begin{equation}\label{CASIMIR}
 C^{(2)}  = \frac{1}{3} (\lambda^2 + \mu^2 + \lambda \mu
+ 3 \lambda + 3 \mu )
\end{equation}
and   
\begin{equation}\label{AUTOG}
 C^{(3)}  = \frac{1}{18} (\lambda - \mu)
(2 \lambda + \mu + 3)(\lambda + 2 \mu + 3)~.
\end{equation}
(Note that an extra factor of 1/2 is needed in Ref.  \cite{Stancu:1991rc} for the eigenvalue of the cubic invariant.)
Then for a $q^3$ system the expectation value of (\ref{QQQ}) is
$\frac{10}{9}$ for a singlet ($\lambda \mu$) = (00)
and $-\frac{5}{36}$ for an octet ($\lambda \mu$) = (11).  
The expectation values of (\ref{QQQBAR}),
necessary in the study of a $q^4 {\overline q}$ system, will be given
in the next section.
    
This is an exploratory study where 
a schematic form for the radial part of
${\mathcal V}_{ijk}$ of (\ref{3BTOTAL}) is assumed.
As in Refs. \cite{Dmitrasinovic:2001nu,Papp:2002xh,Pepin:2001is} we introduce a parameter $c$ representing
the  strength of the three-body relative to the two-body interaction. 
For a $q^3$ system we assume that
\begin{equation}\label{RAD2}
\langle {\mathcal V}_{ijk} \rangle =
c (\langle V_{ij} \rangle + \langle V_{jk} \rangle +
\langle V_{ki} \rangle )~.
\end{equation} 
This is consistent with a triangular shape for a three-body interaction
in baryons.
With such a form the contribution of $V_{2b}$ and $V_{3b}$
add up together and the expectation value of the color part of this sum is
\begin{equation}\label{CHI1}
\chi_1 = \frac{5}{3} + \frac{10}{9} c 
\end{equation}
for a color singlet $q^3$ system \cite{Dmitrasinovic:2001nu,Papp:2002xh,Pepin:2001is}.
To avoid multiple counting, in a $q^4 {\overline q}$ system
we take
\begin{equation}\label{RAD3}
\langle {\mathcal V}_{ijk} \rangle = 
\frac{c}{3} (\langle V_{ij} \rangle + \langle V_{jk} \rangle +
\langle V_{ki} \rangle )~,
\end{equation} 
because each two-body interaction contributes three times.
These assumptions will be used in Sec. IV. 
Next we discuss the basis states.
\vspace{0.5cm}

\section{The basis states}

For describing $q^4 {\overline q}$ systems one can introduce
various coupling schemes, each being convenient for a particular form of the 
interaction operator. For our discussion suppose that the particles
1, 2, 3 and 4 are quarks and 5 is an antiquark.
Then one can first couple three quarks together and then couple this subsystem
to a $q {\overline q}$ pair. In this way one introduces the
so called asymptotic channels having a
physical color singlet - color singlet state 
$|(123)_1 (4 {\overline 5})_1 \rangle$
and two unphysical  
color octet - color octet states $|(123)_8 (4 {\overline 5})_8 \rangle$. 
This coupling
scheme is useful to calculate  matrix elements of the operator (\ref{3Q})
and it gives the appropriate basis at finite distances. At very large distances  only the 
singlet-singlet state  survives, the octet-octet states being
pushed up by the quark-quark interaction. They were first called hidden-color states
in the context of the nucleon-nucleon problem described as a six quark system \cite{Harvey:1988nk}.

Asymptotic channels are convenient to be used  
for all multiquark systems. For example the tetraquarks have one 
singlet-singlet and one octet-octet asymptotic channel \cite{Stancu:1991rc,Brink:1994ic,Stancu:2009ka}.
The hexaquarks have one octet-octet asymptotic channel when orbital excitations 
of permutation symmetry $[42]$ are included \cite{Pepin:2001is}.

The pentaquark system has the particularity that it has two octet-octet
channels, as shown in Appendix \ref{transform}. Thus the asymptotic channels are
\begin{eqnarray}\label{ASYMPT}
|1 \rangle  = |(123)_1~ (4 {\overline 5})_1 \rangle \, \nonumber \\
|2 \rangle  = |(123)^{\rho}_8~ (4 {\overline 5})_8 \rangle \, \nonumber \\
|3 \rangle  = |(123)^{\lambda}_8~ (4 {\overline 5})_8 \rangle . \nonumber \\
\end{eqnarray}
where the $\rho$ and $\lambda$ superscripts, traditionally used for baryons, correspond to the two linear 
independent basis vectors of the mixed irreducible representation $[21]$ of the permutation group S$_3$.
These are basis vectors where the $\rho$ index indicates that the pair 12 is in an antisymmetric state
and the $\lambda$ index indicates that the pair 12 is in symmetric state.

On the other hand  to estimate the contribution of the three-body interaction
(\ref{2QQBAR}) it is convenient first to couple two quarks, say 1 and 2 to the 
antiquark ${\overline 5}$
and then to the subsystem of the remaining pair of quarks, 3 and 4,
to get again total color singlets.
One can construct   
the  following normalized independent states 
\begin{eqnarray}\label{IC}
\,|[{(12)}^S\, {\overline 5}]_{[211]} (34)_{[11] }\rangle \, \nonumber \\ 
\,|[{(12)}^A\, {\overline 5}]_{[211]} (34)_{[11] }\rangle \, \nonumber \\
\,|[{(12)}^A\, {\overline 5}]_{[22]} (34)_{[2] }   \rangle ~.
\end{eqnarray}
The upper index $S (A)$ indicates that the pair 12 is in a symmetric (antisymmetric) state.
Then, the coupling of this pair to the antiquark ${\overline 5}$ leads to three possible color states.
The first two contain a triplet SU(3) $q^2 {\overline q}$ state denoted 
by $[211]$
and the third contains an  antisextet SU(3) state denoted by $[22]$.
Their forms are explicitly given in Appendix  \ref{transform}. 
Other coupling schemes giving rise to a complete set of independent $q^4 \bar q$ color 
singlets are also possible \cite{Park:2017jbn}. 

Of course, the states between different coupling schemes are related to each 
other. In the present case,
we found that the asymptotic  channels (\ref{ASYMPT}) are related to the 
intermediate coupling channels (\ref{IC}) by the following unitary
transformation
\begin{widetext}
\begin{equation}\label{UT}
\renewcommand{\arraystretch}{2.5}\begin{array}{c|ccc}
 & |{ [{(12)}^A\, {\overline 5}]}_{[211]} (34)_{[11] } \rangle  &
   |{ [{(12)}^A\, {\overline 5}]}_{[22]} (34)_{[2] }   \rangle  &
   |{ [{(12)}^S\, {\overline 5}]}_{[211]} (34)_{[11] } \rangle  \\ 
\hline
|({123})_1 ({4 {\overline 5}})_1 \rangle & \sqrt{\frac{1}{3}} &
\sqrt{\frac{2}{3}} & 0 \\
|({123})^{\rho}_8 ({4 {\overline 5}})_8 \rangle & - \sqrt{\frac{2}{3}} &
 \sqrt{\frac{1}{3}} & 0 \\
|({123})^{\lambda}_8 ({4 {\overline 5}})_8 \rangle &  0 &
0 & 1
\end{array}
\end{equation}
\end{widetext}
derived in Appendix \ref{transform}.
The first two rows give transformation coefficients identical to those found 
for tetraquark systems \cite{Stancu:1991rc}. This means that from permutation symmetry 
point of view the structure of the 
corresponding asymptotic basis vectors is the same
in both cases. However  the state 
$|{ [{(12)}^S\, {\overline 5}]}_{[211]} (34)_{[11] } \rangle $
does not exist in tetraquark systems, being incompatible with the definition of an
antiquark as an antisymmetric $qq$ pair. Thus there is only one octet-octet
state in tetraquarks, as it was mentioned above.

The transformation (\ref{UT}) is used in the following to calculate matrix elements of 
the three-body confining interaction potential $V_{3b}$. 
\section{Results}

Here we first calculate the matrix elements of the color part of  
$V_{2b}$ and $V_{3b}$ and then discuss their contribution 
to the total energy of a $q^4 {\overline q}$ system.
The diagonal matrix elements of the two-body interaction $(\ref{V2b})$ 
are calculated from 
\begin{eqnarray}\label{twobody}
\langle i |V_{2b}| i \rangle 
= 3 \langle i |\frac{7}{3} + F_1 \cdot F_2| i \rangle~~~~~~~~~~~~~~
~~~ \nonumber \\
+ 3 \langle i |\frac{7}{3} + F_3 \cdot F_4| i \rangle 
+ 4 \langle i |\frac{7}{3} + F_4 \cdot F_{\overline 5}| i \rangle~.
\end{eqnarray}
where $i,j$ = 1, 2 and 3 are the asymptotic states (\ref{ASYMPT}).
From Appendix \ref{color} one has 
\begin{equation}\label{FdotF}
\langle F_i\cdot F_j \rangle =
\langle F_i\cdot F_{\overline j} \rangle =  - \frac{1}{3}
\end{equation}
both for the singlet-singlet color and the octet-octet color channels.
For 6 pairs of quarks plus 4 pairs quark-antiquark  
the expectation value of the color two-body operator 
becomes  - 10/3  but adding the shift of 7/3  from the definition (\ref{V2b}) or (\ref{twobody})
times 10 one obtains a total of 20 for every asymptotic channel. 
Therefore the integration in the color space of the two-body interaction (\ref{V2b})
for any of the asymptotic states (\ref{ASYMPT}) gives
\begin{equation}
\langle V_{2b}\rangle = 20 \sum\limits_{i<j} V_{ij}.
\end{equation}
The off-diagonal matrix elements of $V_{2b}$ are vanishing for all asymptotic states (\ref{ASYMPT}),
as shown in Appendix \ref{color}.

The matrix elements of the three-body interaction (\ref{3BTOTAL}) in the basis (\ref{ASYMPT})
can be written as
\begin{equation}\label{ME}
\langle  i|V_{3b}|j \rangle = 4 \langle i|{\mathcal C}_{123}|j \rangle
+ 6 \langle i|{\mathcal C}_{12 {\overline 5}}|j \rangle
\end{equation}
where ${\mathcal C}_{123}$ and  ${\mathcal C}_{12 {\overline 5}}$
are defined by (\ref{QQQ}) and (\ref{QQQBAR}) respectively. 
The matrix elements of ${\mathcal C}_{123}$
are straightforwardly obtained 
from the expectation values indicated below Eq. (\ref{AUTOG}).
Using Table \ref{Table1} one can see that the contribution of the 
operator ${\mathcal C}_{12 {\overline 5}}$ vanishes for the color singlet-color singlet state $|1 \rangle$.
The nonvanishing part is due to four times the contribution of ${\mathcal C}_{123}$ which is 
four times the contribution of a $q^3$ system.


\begin{table}
\begin{ruledtabular}
\caption{Expectation values of the operator (\ref{QQQBAR}). \label{Table1} }
\begin{tabular}{cccc}
$state$ & $[(12)^S~{\overline 5}]_{[211]}$  
& $[(12)^A~{\overline 5}]_{[211]}$   
& $[(12)^A~{\overline 5}]_{[22]}$ \vspace{0.2cm}\\
\hline
${\mathcal C}_{ij {\overline k}}$ &  $\frac{5}{18}$ & - $\frac{5}{9}$ 
& $\frac{5}{18}$\vspace{0.2cm} \\
\end{tabular}
\end{ruledtabular}
\end{table}

To calculate all the other matrix elements of ${\mathcal C}_{12 {\overline 5}}$
we make use of the unitary transformation (\ref{UT}) which expresses
the asymptotic channels as linear combination of the 
basis vectors (\ref{IC}) and  then use 
the expectation values  given in Table \ref{Table1}. 
The Ansatz (\ref{RAD3}) allows to introduce a common 
radial factor for $V_{2b}$ and $V_{3b}$, namely $\sum\limits_{i<j} V_{ij}$.
In this way
we have obtained the color part of $V_{2b}+V_{3b}$ as given by the matrix
\begin{equation}\label{MATRIX}
\renewcommand{\arraystretch}{2.5}\begin{array}{c|ccc}
 &  \langle 1|1 \rangle  & \langle 2|2 \rangle  &  \langle 3|3 \rangle \\
\hline
\langle 1|1 \rangle  & 20 + \frac{40}{9}~c &
 \frac{10\sqrt{2}}{6}~c & 0 \\
\langle 2|2 \rangle & \frac{10\sqrt{2}}{6}~c &
 20 - \frac{20}{9}~c & 0 \\
\langle 3|3 \rangle & 0 &
0 & 20 + \frac{10}{9}c
\end{array}
\end{equation}
where $i$ = 1, 2, 3 are the asymptotic states (\ref{ASYMPT}).
The eigenvalues of this matrix are
\begin{equation}\label{EIGEN}
e_{1,2} = 20 + \frac{10}{9} c \pm \frac{10}{\sqrt{6}} c ,~~~~~ \, \quad \\
e_3 = 20 + \frac{10}{9} c~.
\end{equation}
To get the full contribution of the confinement one must
multiply each eigenvalue by $\sum\limits_{i<j} V_{ij}$.
The eigenvector associated with $e_1$ is dominantly color singlet - color 
singlet
(the state $|1 \rangle $ appears with 91 \% probability) and the 
eigenvector associated with $e_2$ is dominantly
color octet - color octet (the state $|2 \rangle $ appears
with 91 \% probability) irrespective of the value of $c$.
The eigenvector $e_3$ is a pure color octet-color octet state as it can
be seen from Eq. (\ref{MATRIX}). It is stable against strong decay into
a baryon plus a meson.

In  Ref. \cite{Dmitrasinovic:2001nu} the range proposed for the relative strength $c$ was
\begin{equation}\label{C}
- \frac{3}{2} < c < \frac{2}{5}~.
\end{equation}
The upper limit ensures that the lowest color singlet appears below the 
lower color octet in a $q^3$ system and the lower limit is required by 
the stability condition of the nucleon, $\langle V_{2b} + V_{3b} \rangle > 0.$
In Ref. \cite{Pepin:2001is} some arbitrariness was noticed regarding the lower limit.
For our discussion it is enough to restrict the interval to
\begin{equation}\label{CNEW}
- 1.0 \leq c < 0.4~.
\end{equation}

In the stability problem against strong decays of a $q^4 {\overline q}$ system we are 
interested in the quantity
\begin{equation}
\Delta E = E(q^4 {\overline q}) - E(q^3) - E(q {\overline q}) 
\end{equation}
where $E(q^4 {\overline q})$ represents the lowest energy of the 
$q^4 {\overline q}$ system  and $E(q^3) + E(q {\overline q})$
is the threshold energy for the decay of $q^4 {\overline q}$
into a baryon described as a $q^3$ system and a meson $q {\overline q}$.
The condition  $\Delta E < 0 $ is interpreted as stability, otherwise
the system is unstable against strong decays.
Of course the  Hamiltonian (\ref{HAMILT}) must also be used in the
calculation of $E(q^3)$. In each case a hyperfine interaction should be
added but this is beyond the scope of the present study.
As mentioned in Sec. II the contribution of $V_{2b}+ V_{3b} $
to a color singlet $q^3$ state is proportional to $\chi_1$ 
given by (\ref{CHI1}).
In order to keep the ground state $E(q^3)$ unchanged in the 
presence of the three-body force we have to rescale $E(q^3)$ by  dividing it
by $\chi_1$ as in Ref. \cite{Papp:2002xh}. For consistency one must also
divide the eigenvalues (\ref{EIGEN}) by the same quantity. 

\begin{figure}
\epsfysize=5.5cm \epsffile{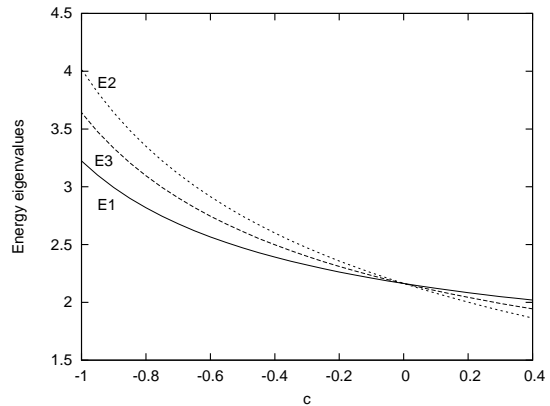}  
\caption{\small Eigenvalues $E_i$ (GeV)
given by Eq. (\ref{HO}) as a function of the relative
strength $c$ of the three-body relative to the two-body confinement. }
\label{fig1}
\end{figure}

It is useful to make an estimate for a harmonic oscillator confinement.
For five quarks (antiquarks) of equal masses the eigenvalues of
(\ref{HAMILT}) are  \cite{Stancu:1997dq}
\begin{equation}
E_i=(N + 6)\hbar \omega_i
\end{equation}
where $N$ is the number of quanta and we take $N=0$. The frequency 
$\omega_i$ of a five body system is related to the frequency $\omega_0$
of a three body system by \cite{Helminen:2000jb}
\begin{equation}
\omega_i = \sqrt{5/6}~ \omega_0 .
\end{equation}
We normalize the eigenvalues $E_i$ 
of (\ref{HAMILT}) 
such as in the absence
of three-body forces ($c$=0) to obtain this relation.
In addition, as we deal with 10 pairs of quarks we have to divide each eigenvalue of $H$ by 10
to be consistent with Ref. \cite{Papp:2002xh}.
Then,  in terms of $e_i$ of Eqs. (\ref{EIGEN}) we have
\begin{equation}\label{HO}
E_i = 6 \hbar \omega_i = 5 \sqrt{\frac{e_i}{10 \chi_1}} \hbar \omega_0~,
\end{equation}
where $\chi_1$ is defined by Eq. (\ref{CHI1}).
For $\hbar \omega_0$ = 395 MeV \cite{Papp:2002xh}
the three resulting eigenvalues are displayed
in Fig. \ref{fig1}.
Without a three-body force ($c$ = 0) the states are degenerate.
The three-body confining force ($c \neq 0$) introduces a splitting. A negative 
$c$ increases all  $E_i$ with respect to the values taken at 
$c$ = 0. For example one has $E_1(c=-1)-E_1(c=0)$= 1.06 GeV.
This means that the pentaquarks become less 
and less stable with a decreasing  negative $c$.
The lowest eigenvalue is $E_1$ which is 
dominantly color singlet $q^3$ color singlet $q {\overline q}$.
For $c>0$  the lowest state becomes $E_2$ which is dominantly
color octet - color octet. 
One has $E_2(c=0)-E_2(c=0.35)$ = 267 MeV, which is a substantial lowering.

Thus the three-body force 
helps now in stabilizing the $q^4 {\overline q}$ system against strong decays  
because $E_2$ is dominantly  a color 
octet - color octet state thus it has a small amplitude to decay strongly.  
However the gap between $E_2$ and the other states is not so large. 
For example, $E_2(c=0.35)-E_1(c=0.35)$= - 138 MeV.    
This means that through the addition of a hyperfine interaction to
the Hamiltonian (\ref{HAMILT}) 
the proportion of singlet-singlet and octet-octet contributions might
change.

The $c > 0$ case also brings a new perspective to baryon spectroscopy.
If the energy of the $q^4 {\overline q}$ states are close to those of
the $q^3$ states one should expect significant admixtures of 
 $q^4 {\overline q}$ states in
the baryon resonance wave functions. The low lying  $q^4 {\overline q}$ states
provide an 
approximation to the open baryon-meson channels that 
give rise to large strong decay widths of the empirical resonances 
\cite{Helminen:2000jb}.
Let us give an estimate for energies relevant for the Roper resonance.
In constituent quark models this is usually described as a 2 $\hbar \omega_0$
excitation. In these models the presence of a negative additional constant
$V_0$ in the Hamiltonian is also required for a realistic description of the
spectrum. 
Accordingly, if we define the "unperturbed" energy 
(no hyperfine interaction) of a $q^3$ system as 
$E(q^3)= 3m + 3V_0 + 5\hbar \omega_0$  with $m=340$ MeV and $V_0$= - 269 MeV
as in  Ref. \cite{Helminen:2000jb} and use, as above, $\hbar \omega_0$ = 395 MeV
we obtain $E(q^3)$ = 2188 MeV. If  we estimate the 
energy of the lowest $q^4 {\overline q}$ in a similar manner we have
$E(q^4 {\overline q}) = 5m + 5 V_0 +E_2$. At $c=0.35$ one has $E_2$= 1897 Mev from
which one obtains $E (q^4 {\overline q})$ = 2252 MeV, very close to 
$E(q^3)$. Thus the 
three-body confining interaction with a positive $c$ could provide a
mechanism to lower the $q^4 {\overline q}$ states, which is an alternative
to the mechanism proposed in Ref. \cite{Helminen:2000jb} based on a 
schematic flavor and spin dependent interaction where the strength of this 
interaction is adjusted to allow admixtures.

\section{Conclusions}

Here we have shown that the three-body confining interaction 
(\ref{3BTOTAL})-(\ref{2QQBAR})
with a negative $c$ destabilizes the pentaquarks
while a positive $c$ helps in stabilizing them. 
A similar conclusion has been drawn in Ref. \cite{Pepin:2001is} 
in connection with tetraquarks.
The conclusion also holds for tetraquarks or 
pentaquarks containing heavy flavors, inasmuch as
the confinement interaction is flavor independent.

The unitary transformation (\ref{UT}) derived here relates two complete sets 
of color singlet bases needed to describe $q^4 {\overline q}$ systems. Each contain
three independent states. The set (\ref{ASYMPT}) is the most suitable 
to study the stability against strong decays. 
The set (\ref{IC}) may be useful in diquark-triquark type models \cite{Karliner:2003dt}.
Thus 
in a $q^4 {\overline q}$ system there are three asymptotic
channels in the color space as compared to two for $q^2 {\overline q}^2$ 
systems. 

As mentioned in the introduction the stability of multiquark systems depends on the 
constituent masses as well. In particular, for unequal masses the kinetic term deserves a 
special attention, by studying the role of both its symmetric and asymmetric parts,
as it has been done for tetraquarks in Ref. \cite{Richard:2017vry}. We are aware that 
the problem becomes complex  if the pentaquark contains more than two flavors, as for example, 
those proposed in Refs. \cite{Gignoux:1987cn,Lipkin:1987sk}.

Similar to  tetraquarks \cite{Brink:1994ic},
one can construct full wave functions including
the position, the spin and the flavor degrees of freedom  to
calculate  the total energy of  $q^4 {\overline q}$ 
by adding a hyperfine interaction to see which is the role of
octet-octet states in the stability. In this way one may perhaps avoid
the fall apart decay mode of the ground state pentaquarks obtained
in Ref.  \cite{Park:2017jbn}. 

It would certainly be interesting to make contact with the many-quark confining 
forces from flux-tube models inspired by the strong coupling limit of QCD as 
proposed in Ref. \cite{Vijande:2007ix} for tetraquarks and extended to 
pentaquarks in Ref. \cite{Richard:2009rp}. For pentaquarks two distinct 
contributions should be considered, as depicted in Fig. 10 of Ref. \cite{Richard:2016eis}.
These are a disconnected and a connected flux-tube diagrams, the first corresponding
to the baryon-meson configuration and the second to a Steiner tree diagram \cite{Richard:2009rp}. 
In a color basis, the first should be a color singlet-color singlet state and the 
second should be related to two distinct color triplet - color antitriplet states. These must be associated 
to a cluster of four quarks and the antitriplet must be associated to the antiquark. 
One has to establish the relation of this basis to that of Eq. (\ref{ASYMPT})
each involving  different intermediate couplings. 
Thus more algebraic work 
is needed to make a link between flux-tube confining potentials and the present study. 
If successful, a linear radial dependence, as supported by lattice studies
for the three-body confining potential, could be considered in the future.

\appendix

\section{The SU(3) antiquark generators}\label{fbar}
We find it useful to exhibit the action of the 
antiquark generators, Eq. (\ref{Fantiquark}), 
\begin{equation}
{\overline F}^a = -\frac{1}{2} \lambda^{a*}, (a = 1,...,8)
\end{equation}
on antiquark states $\bar q_i$, needed in Appendix \ref{alternative}. 
These are
\begin{eqnarray}\label{myeqs}
{\overline F}^1 {\overline q}_1  = - \frac{1}{2} {\overline q}_2,~~~~~ {\overline F}^1  {\overline q_2} = - \frac{1}{2} {\overline q_1}, 
 ~~~~~ {\overline F}^1  {\overline q_3} = 0, \\
{\overline F}^2  \bar q_1 =  \frac{i}{2} \bar q_2,~~~~~ {\overline F}^2  \bar q_2 = -  \frac{i}{2}\bar q_1, 
 ~~~~~ {\overline F}^2  \bar q_3 = 0, \\
{\overline F}^3  \bar q_1 = - \frac{1}{2} \bar q_1,~~~~~ {\overline F}^3  \bar q_2 =   \frac{1}{2}\bar q_2, 
 ~~~~~ {\overline F}^3  \bar q_3 = 0, \\
{\overline F}^4  \bar q_1 = - \frac{1}{2} \bar q_3,~~~~~ {\overline F}^4  \bar q_2 = 0, 
 ~~~~~ {\overline F}^4  \bar q_3 = - \frac{1}{2} \bar q_1, \\
{\overline F}^5  \bar q_1 =  \frac{i}{2} \bar q_3,~~~~~ {\overline F}^5  \bar q_2 = 0, 
 ~~~~~ {\overline F}^5  \bar q_3 = - \frac{i}{2} \bar q_1, \\
{\overline F}^6  \bar q_1 = 0,~~~~~ {\overline F}^6  \bar q_2 = - \frac{1}{2} \bar q_3, 
 ~~~~~ {\overline F}^6  \bar q_3 = - \frac{1}{2} \bar q_2, \\
{\overline F}^7  \bar q_1 = 0,~~~~~ {\overline F}^7  \bar q_2 = \frac{i}{2} \bar q_3, 
 ~~~~~ {\overline F}^7  \bar q_3 = - \frac{i}{2} \bar q_2, \\
{\overline F}^8  \bar q_1 = -\frac{1}{2\sqrt{3}} \bar q_1  ,~~~~~ {\overline F}^8  \bar q_2 = -\frac{1}{2\sqrt{3}} \bar q_2, \nonumber \\ 
 ~~~~~ {\overline F}^8  \bar q_3 =  \frac{1}{\sqrt{3}} \bar q_3.
\end{eqnarray}

In the flavor space one can identify  
\begin{equation}
{\overline q}_1 = \bar u, ~~~{\overline q}_2 = \bar d, ~~~{\overline q}_3 =  \bar s. 
\end{equation}
As a check, using the above relations and the symmetric constants $d^{abc}$ from Table \ref{detail} one 
can obtain the eigenvalue of the cubic invariant operator for antiquarks,
in the Lorentz vector model \cite{Dmitrasinovic:2001nu}
\begin{equation}\label{3BARS}
{ \overline {\mathcal C}}_{ijk} =  d^{abc} ~{\overline F}^{a}_i
~{\overline F}^{b}_j ~ {\overline F}^{c}_k
\end{equation}
by acting on $\bar d$, the highest weight state of the antitriplet $(\lambda\mu)$ = (01).
The result is
\begin{equation}
{\overline C}^{(3)} \bar d = - \frac{10}{9} \bar d
\end{equation} 
which means that the eigenvalue of the cubic invariant has an opposite sign to that of the triplet $u,d,s$,
having   $(\lambda,\mu)$ = (10),  consistent with Eq. (\ref{AUTOG}).

\section{Matrix elements for  two-body operators}\label{color}
Here we prove Eq. (\ref{FdotF}) in a simplified way. This is a more explicit proof
of the relation derived in Ref. \cite{Genovese:1997tm}.
We first need to 
evaluate the color operator matrix element of a subsystem of four quarks 
\begin{equation}
\langle O^C \rangle = \langle\sum_{i<j}^{4} F_i \cdot F_j \rangle
= \frac{1}{2}(C^{(2)} - 4 C^{(2)}_q) , 
\label{colour}
\end{equation}
where $C^{(2)}$ is the Casimir operator eigenvalue (\ref{CASIMIR}) for a system of four quarks and
$C^{(2)}_q = 4/3$ is the eigenvalue of the Casimir operator for
a quark, light or heavy. The color state of the four-quark subsystem is $[211]_C$ so that one has $C^{(2)} = 4/3$. Hence
$\langle O^C \rangle_{[211]} = -2$.  As there are 6 pairs in (\ref{colour}),
each pair of quarks will have
\begin{equation}
\langle F_i \cdot F_j \rangle = -\frac{1}{3}.
\end{equation}
For the antiquark the Casimir operator eigenvalue is $C^{(2)}_{\bar{q}}
= 4/3$, where $\bar{q}$ can be light or heavy.  Then for a pentaquark, in a color-singlet state
$[222]$, formed of a subsystem of four quarks and an antiquark,  
one can define the color operator eigenvalue
\begin{equation}
\langle\sum_{i=1}^{4} F_i \cdot F_{\bar{5}} \rangle = (C^{(2)}_{[222]} - C^{(2)}_q - C^{(2)}_{\bar{q}})/2 
= - \frac{4}{3},
\end{equation}
where we have used the Casimir operator eigenvalue $C^{(2)}_{[222]}$ = 0 of the whole system in a color singlet,
and $C^{(2)}_q$ = $C^{(2)}_{\bar{q}}$ = 4/3. 
As each quark shares 1/4 one can conclude that in average, for every asymptotic channel  (\ref{ASYMPT})
we have
\begin{equation}
\langle  F_i \cdot F_j   \rangle 
= - \frac{1}{3}, \hspace{0.5cm} i= 1,2,3,4, \, \hspace{0.2cm} j=\bar{5} ,
\end{equation}
which means that the color expectation value for any pair, quark-quark or a quark-antiquark
 is $-1/3$. There are in all 10 pairs
so that for a two-body color confinement with a constant $V_{ij}$, the color operator gives
\begin{equation}
\sum\limits_{i<j} \langle F_i \cdot F_j \rangle = - \frac{10}{3}.
\end{equation}

There is another simplified, but slightly more elaborate, way to derive Eq. (\ref{FdotF}).
We introduce the scalar product $F_{123} \cdot F_{4 \bar 5}$ defined as
\begin{equation}
F_{123} \cdot F_{4 \bar 5} = (F_1 + F_2 + F_3) \cdot (F_4 + F_{\bar 5}),
\end{equation}
or alternatively
\begin{equation}
F_{123} \cdot F_{4 \bar 5} = F_1 \cdot F_4 + F_2 \cdot F_4 +  F_3 \cdot F_4 
+ F_1 \cdot F_{\bar 5} +  F_2 \cdot F_{\bar 5} + F_3 \cdot F_{\bar 5}.
\end{equation}  
In terms of the Casimir operators eigenvalues of the whole system $C^{(2)}_{[222]}$ and of the separate subsystems $C^{(2)}_{123}$ and 
$C^{(2)}_{4 \bar 5}$ one has
\begin{equation}\label{product} 
F_{123} \cdot F_{4 \bar 5} = \frac{1}{2}(C^{(2)}_{[222]} - C^{(2)}_{123} - C^{(2)}_{4 \bar 5}).
\end{equation} 
For the asymptotic channel $|1\rangle$ we get 
\begin{equation}
\langle 1 |F_{123} \cdot F_{4 \bar 5}|1 \rangle = 0.
\end{equation} 
On the other hand it is reasonable to assume that 
\begin{eqnarray} 
\langle F_1 \cdot F_4 \rangle = \langle F_2 \cdot F_4 \rangle =  \langle F_3 \cdot F_4 \rangle,  \nonumber\\
\langle F_1 \cdot F_{\bar 5} \rangle = \langle F_2 \cdot F_{\bar 5} \rangle = \langle F_3 \cdot F_{\bar 5} \rangle,
\end{eqnarray} 
so that we have 
\begin{equation}\label{rel}
\langle F_i \cdot F_4 \rangle + \langle F_i \cdot F_{\bar 5} \rangle = 0, \hspace{1cm} i = 1,2,3.
\end{equation}
The nonvanishing terms give 
\begin{eqnarray}
\langle 1 |\sum_{i<j}  F_i \cdot F_j | 1 \rangle = \langle 1| F_1 \cdot F_2 | 1 \rangle + \langle 1| F_1 \cdot F_3 | 1 \rangle \nonumber \\
+ \langle 1| F_2 \cdot F_3 | 1 \rangle + \langle 1| F_4 \cdot F_{\bar 5} | 1 \rangle.
\end{eqnarray}
As we know 
\begin{equation}
\langle 1| F_i \cdot F_j | 1 \rangle = - \frac{2}{3}, \hspace{0.5cm} \langle 1| F_4 \cdot F_{\bar 5} | 1 \rangle = - \frac{4}{3},
\end{equation}
we obtain 
\begin{equation}
\langle 1 |\sum_{i<j}  F_i \cdot F_j | 1 \rangle = - \frac{10}{3},
\end{equation}
as above. 
For the hidden-color (octet-octet) states $|2 \rangle $ and $|3 \rangle $ 
Eq. (\ref{product})  gives
\begin{equation}
\langle (123)_8 (4 {\overline 5})_8|   F_{123} \cdot F_{4 \bar 5}  |(123)_8 (4 {\overline 5})_8 \rangle = - 3,
\end{equation}
from where the analog of the relation (\ref{rel}) becomes 
\begin{equation}
\langle F_i \cdot F_4 \rangle + \langle F_i \cdot F_{\bar 5} \rangle = - 1. \hspace{1cm} i = 1,2,3.
\end{equation}
Then, for a color octet three quark states we take the average between symmetric and antisymmetric states
of particles 1 and 2 to get 
\begin{equation}
\langle F_i \cdot F_j \rangle = \frac{1}{2}(\frac{1}{3} - \frac{2}{3}) = - \frac{1}{6}, \hspace{1cm} i = 1,2,3.
\end{equation}
Then it follows that for any of the hidden color states one has
\begin{eqnarray}
\langle (123)_8 (4 {\overline 5})_8| \sum_{i<j}  F_i \cdot F_j |\langle (123)_8 (4 {\overline 5})_8 \rangle = 3 \langle F_1 \cdot F_2 \rangle 
\nonumber \\
 + 3 (\langle F_1 \cdot F_4 \rangle + \langle F_1 \cdot F_{\bar 5} \rangle) 
+ \langle F_4 \cdot F_{\bar 5} \rangle  \nonumber \\ 
= 3  ( - \frac{1}{6}) - 3 + \frac{1}{6} =  - \frac{10}{3}\nonumber \\ 
\end{eqnarray}
which is the same as for the singlet-singlet state. 

This proof is convenient for our purpose where the confinement is a constant 
in the configuration space. But if a radial shape is introduced then one has to
proceed as in the appendix of Ref. \cite{Park:2017jbn}, which gives a consistent result 
with ours.

Using the explicit form of the color basis states $|1 \rangle$, $|2 \rangle$ and  $|3 \rangle$ of Appendix \ref{transform}
one can calculate the required off-diagonal matrix elements of $V_{2b}$. For quark-quark pairs they are
\begin{eqnarray}
\langle 1 | F_1 \cdot F_2 | 2 \rangle = \langle 1 | F_1 \cdot F_3 | 2 \rangle = \langle 1 | F_2 \cdot F_3 | 2 \rangle = 0, \nonumber\\
\langle 1 | F_1 \cdot F_4 | 2 \rangle = \langle 1 | F_2 \cdot F_4 | 2 \rangle = - \frac{\sqrt{2}}{6}, \nonumber\\
\langle 1 | F_3 \cdot F_4 | 2 \rangle = \frac{\sqrt{2}}{3},
\end{eqnarray}
and for quark-antiquark pairs one has
\begin{eqnarray}
\langle 1 | F_1 \cdot F_{\bar 5} | 2 \rangle = \langle 1 | F_2 \cdot F_{\bar 5} | 2 \rangle = \frac{1}{\sqrt{18}}, \nonumber\\
\langle 1 | F_3 \cdot F_{\bar 5} | 2 \rangle = - \frac{2}{\sqrt{18}}, ~~~ \langle 1 | F_4 \cdot F_{\bar 5} | 2 \rangle = 0.
\end{eqnarray}
This implies that 
\begin{equation}
\sum_{i < j}^4 \langle 1 | F_i \cdot F_j | 2 \rangle  + \sum_i^4 \langle 1 | F_i \cdot F_{\bar 5} | 2 \rangle = 0
\end{equation}
which proves that the off-diagonal matrix element of $V_{2b}$ between states $|1 \rangle$ and $|2 \rangle$ is zero.
Similarly one can prove that the other off-diagonal matrix element of $V_{2b}$ are vanishing.
For this  purpose one needs to know that
\begin{eqnarray}
\langle 1 | F_1 \cdot F_2 | 3 \rangle = \langle 1 | F_1 \cdot F_3 | 3 \rangle = \langle 1 | F_2 \cdot F_3 | 3 \rangle = 0, \nonumber\\
\langle 1 | F_1 \cdot F_4 | 3 \rangle = - \langle 1 | F_2 \cdot F_4 | 3 \rangle =  \frac{1}{\sqrt{6}}, \nonumber\\
\langle 1 | F_3 \cdot F_4 | 3 \rangle = 0,
\end{eqnarray}
and
\begin{eqnarray}
\langle 1 | F_1 \cdot F_{\bar 5} | 3 \rangle = - \langle 1 | F_2 \cdot F_{\bar 5} | 3 \rangle = - \frac{1}{\sqrt{6}}, \nonumber\\
\langle 1 | F_3 \cdot F_{\bar 5} | 3 \rangle = 0, ~~~ \langle 1 | F_4 \cdot F_{\bar 5} | 3 \rangle = 0.
\end{eqnarray}

\section{Two useful color singlet bases for pentaquarks}\label{transform}

We have used the tensor method \cite{Stancu:1991rc} to write down explicit expressions for
the normalized basis states appearing in the unitary transformation (\ref{UT})
in terms 
of their quark content. By equalizing the coefficients
of identical terms of both sides we have obtained a system of linear
equations for the matrix elements of the unitary transformation.

Let us denote by $q_i$ and ${\overline q}_i$  ($i = $ 1,2 and 3) the three 
possible color states of a quark and an antiquark respectively. 
By using the tensor method \cite{Stancu:1991rc} the 
asymptotic channel states can be easily constructed as scalars\cite{Richard:2009rp}
from the product between a flavor singlet (flavor octet)  $q^3$ state and 
a flavor singlet (flavor octet) $q {\overline q}$ state. The three possible
normalized colorless $q^4 {\overline q}$ states are 
\begin{equation}\label{SINGLET}
|1 \rangle = |(123)_1 (4 {\overline 5})_1 \rangle =
\sqrt{\frac{1}{18}} \varepsilon_{\alpha \beta \gamma} 
q^{\alpha} q^{\beta} q^{\gamma} 
~\delta^{\nu}_{\mu} q^{\mu} {\overline q}_{\nu}~, 
\end{equation}
\begin{eqnarray}\label{ANTISYM}
|2 \rangle = |(123)^{\rho}_8 (4 {\overline 5})_8 \rangle =
\frac{1}{12} \times ~~~~~~~~~~~~~~~~~~ \nonumber \\
\{  3(q^1 q^2 q^1 - q^2 q^1 q^1) q^3 {\overline q}_1 
+ 3(q^1 q^2 q^2 - q^2 q^1 q^2) q^3 {\overline q}_2 \nonumber \\
- 3(q^1 q^3 q^1 - q^3 q^1 q^1) q^2 {\overline q}_1
+ 3(q^2 q^3 q^2 - q^3 q^2 q^2) q^1 {\overline q}_2 \nonumber \\
- 3(q^1 q^3 q^3 - q^3 q^1 q^3) q^2 {\overline q}_3
+ 3(q^2 q^3 q^3 - q^3 q^2 q^3) q^1 {\overline q}_3  \nonumber \\
-  (q^1 q^2 q^3 - q^2 q^1 q^3 + q^3 q^1 q^2 - q^1 q^3 q^2 
+ 2 q^3 q^2 q^1 \nonumber \\
- 2 q^2 q^3 q^1) q^1 {\overline q}_1 
-  (q^1 q^2 q^3 - q^2 q^1 q^3 - 2 q^3 q^1 q^2 + 2 q^1 q^3 q^2 \nonumber \\
- q^3 q^2 q^1 
+ q^2 q^3 q^1) q^2 {\overline q}_2  
+ (2 q^1 q^2 q^3 - 2 q^2 q^1 q^3 - q^3 q^1 q^2 \nonumber \\ 
+ q^1 q^3 q^2 + q^3 q^2 q^1 - q^2 q^3 q^1) q^3 {\overline q}_3  \} \nonumber \\
\end{eqnarray}
and
\begin{eqnarray}\label{SYM}
|3 \rangle = |(123)^{\lambda}_8 (4 {\overline 5})_8 \rangle =
\sqrt{\frac{1}{48}} \times ~~~~~~~~~~~~~~~~~~ \nonumber \\
\{ - (q^1 q^2 q^1 + q^2 q^1 q^1 -2 q^1 q^1 q^2) q^3 {\overline q}_1\nonumber \\
   + (q^1 q^2 q^2 + q^2 q^1 q^2 -2 q^2 q^2 q^1) q^3 {\overline q}_2\nonumber \\
   + (q^1 q^3 q^1 + q^3 q^1 q^1 -2 q^1 q^1 q^3) q^2 {\overline q}_1\nonumber \\
   - (q^2 q^3 q^2 + q^3 q^2 q^2 -2 q^2 q^2 q^3) q^1 {\overline q}_2\nonumber \\
   - (q^1 q^3 q^3 + q^3 q^1 q^3 -2 q^3 q^3 q^1) q^2 {\overline q}_3\nonumber \\
   + (q^2 q^3 q^3 + q^3 q^2 q^3 -2 q^3 q^3 q^2) q^1 {\overline q}_3\nonumber \\
+(q^1 q^2 q^3 + q^2 q^1 q^3 -  q^3 q^1 q^2 -  q^1 q^3 q^2)
q^1 {\overline q}_1 \nonumber \\
-(q^1 q^2 q^3 + q^2 q^1 q^3 -  q^3 q^2 q^1 -  q^2 q^3 q^1)
q^2 {\overline q}_2 \nonumber \\
+(q^3 q^1 q^2 + q^1 q^3 q^2 - q^3 q^2 q^1 - q^2 q^3 q^1) q^3 {\overline q}_3 
\} 
\end{eqnarray}
Note that the normal order of particles is understood everywhere.
The difference between (\ref{ANTISYM}) and (\ref{SYM}) is that 
in the first the particles 1 and 2 are in an antisymmetric state 
while in the second they are in a symmetric state. 
That is why they carry the upperscripts ${\rho}$ and  ${\lambda}$
respectively, like the mixed symmetry $q^3$ states used in baryon
spectroscopy.

For the basis vectors in the intermediate coupling
we use the tensor method first to construct $qq {\overline q}$ states.

1) The three components  
of $[(12)^S {\overline 5}]_{[211]}^i$ state are obtained from by contraction 
with $\delta^k_i$
\begin{equation} 
T^j = T^{ij}_k {\delta}^k_i
\end{equation}
from the  tensor $T^{ij}_k$ which is symmetric in the
upper indices
\begin{equation}
T^{ij}_k = (q^i q^j + q^j q^i) {\overline q}_k
\end{equation}
The normalized components are
\begin{eqnarray}\label{TWOSYM}
|[{(12)}^S\, {\overline 5}]_{[211]} \rangle^1 = 
\frac{|}{2 \sqrt{2}} [2 q^1 q^1 {\overline q}_1 + \nonumber \\
(q^1 q^2 + q^2 q^1) {\overline q}_2
+(q^3 q^1 + q^1 q^3) {\overline q}_3~. \nonumber \\
|[{(12)}^S\, {\overline 5}]_{[211]} \rangle^2 = 
\frac{|}{2 \sqrt{2}} [2 q^2 q^2 {\overline q}_2 + \nonumber \\
(q^2 q^3 + q^3 q^2) {\overline q}_3
+(q^1 q^2 + q^2 q^1) {\overline q}_1~. \nonumber \\
|[{(12)}^S\, {\overline 5}]_{[211]} \rangle^3 = 
\frac{|}{2 \sqrt{2}} [2 q^3 q^3 {\overline q}_3 + \nonumber \\
(q^3 q^1 + q^1 q^3) {\overline q}_1
+(q^2 q^3 + q^3 q^2) {\overline q}_2~. \nonumber \\
\end{eqnarray}

2) The three components of the state $[(12)^A {\overline 5}]_{[211]}^i$ 
form the antisymmetric tensor 
\begin{equation}
T^i = {\epsilon}^{ijk} T_j T_k  
\end{equation}
In this antisymmetric product the first factor represents a $qq$ subsystem
described by 
\begin{equation}\label{CONTRAV} 
T_j = {\epsilon}_{jlm} q^l q^m
\end{equation}
and the second an antiquark seen as a  $qq$ pair
\begin{eqnarray}\label{ANTIQ} 
{\overline q}_1= \frac{1}{\sqrt{2}}(q^2 q^3 -q^3 q^2)\,, \quad 
{\overline q}_2= \frac{1}{\sqrt{2}}(q^3 q^1 -q^1 q^3)\,, \quad \nonumber \\
{\overline q}_3= \frac{1}{\sqrt{2}}(q^1 q^2 -q^2 q^1).~~~~~~~~~~~~~~~~
\end{eqnarray}

\noindent
Then the first normalized components is
\begin{eqnarray}\label{TWOASYM}
[(12)^A {\overline 5}]_{[211]}^1 =
\frac{1}{2}[(q^3 q^1 - q^1 q^3) {\overline q}_3 
- (q^1 q^2 - q^2 q^1) {\overline q}_2]~. \nonumber \\
\end{eqnarray}
The other two are obtained by circular permutations.

To form a $q^4 {\overline q}$ singlet state together with particles 3 and 4
in the above two cases one must 
construct the scalar product
$\frac{1}{\sqrt{3}} T^i T_i$ where $T^i$ is either of the form (\ref{TWOSYM}) 
or (\ref{TWOASYM}) plus circular permutations. The tensor $T_i$
made of the particles 3 and 4 is defined by (\ref{CONTRAV}).

3) The six components of $[(12)^A {\overline 5}]_{[22]}^i$
are defined by the symmetric tensor
\begin{equation}\label{TSYM}
T_{ij} = \frac{1}{2}(T_i T_j + T_j T_i)
\end{equation}
where the first factor in the right hand side should be replaced by
(\ref{CONTRAV}) and the second by (\ref{ANTIQ}). Then, for example the
first component becomes
\begin{equation}
[(12)^A {\overline 5}]_{[22]}^1 = \frac{1}{2}
\{ (q^3 q^1 - q^1 q^3) {\overline q}_3 + 
(q^1 q^2 - q^2 q^1) {\overline q}_2 \}~.
\end{equation}

To form a $q^4 {\overline q}$ singlet state
one needs the symmetric tensor
\begin{equation} 
T^{ij}= q^i q^j + q^j q^i
\end{equation}
associated to the particles 3 and 4.
Then the $q^4 {\overline q}$ singlet state is defined by the
scalar product 
$\frac{1}{\sqrt{6}} T^{ij} T_{ij}$.

In this way we have obtained all states needed to determine the unitary 
transformation (\ref{UT}).

\section{Matrix elements of ${\mathcal C}_{ijk}$ and ${\mathcal C}_{ij\overline k}$}\label{alternative}

\begin{table}
\caption[colour]{\label{detail} The contributing terms to the matrix elements of
operators defined by Eqs. (\ref{3Q}) and (\ref{2QQBAR}) for the specific cases
$\langle 3 | d^{abc} F^{a}_{1} F^{b}_{2} F^{c}_{4} | 3 \rangle $ and 
$\langle 3 | d^{abc} F^{a}_{1} F^{b}_{2} \overline{F}^{c}_{5} | 3 \rangle $. 
The first column gives the color indices $(abc)$, the
second the corresponding constant $d^{abc}$, the third is the multiplicity of each term and the fourth and fifth columns the value
of the matrix element for a given $d^{abc}$.\\}
\begin{ruledtabular}
\begin{tabular}{c|c@{\hspace{3mm}}|c|c@{\hspace{2mm}}|c@{\hspace{2mm}}}
$(abc)$ & $d^{abc}$ & $m^{abc}$ & $\langle 3 | d^{abc} F^{a}_{1} F^{b}_{2} F^{c}_{4} | 3 \rangle $
 & $\langle 3 | d^{abc} F^{a}_{1} F^{b}_{2} \overline{F}^{c}_{5} | 3 \rangle $ \\
\hline
118 & $\sqrt{3}/3$ & 3 & -1/288 & -1/144 \\
146 &          1/2 & 6 & -1/384 & -1/192 \\
157 &          1/2 & 6 & -1/384 & -1/192 \\
228 & $\sqrt{3}/3$ & 3 & -1/288 & -1/144 \\
247 &         -1/2 & 6 & -1/384 & -1/192 \\
256 &          1/2 & 6 & -1/384 & -1/192 \\
338 & $\sqrt{3}/3$ & 3 & -1/288 & -1/144 \\
344 &          1/2 & 3 & -1/384 & -1/192 \\
355 &          1/2 & 3 & -1/384 & -1/192 \\
366 &         -1/2 & 3 & -1/384 & -1/192 \\
377 &         -1/2 & 3 & -1/384 & -1/192 \\
448 &-$\sqrt{3}/6$ & 3 &-1/1152 & -1/576 \\
558 &-$\sqrt{3}/6$ & 3 &-1/1152 & -1/576 \\
668 &-$\sqrt{3}/6$ & 3 &-1/1152 & -1/576 \\
778 &-$\sqrt{3}/6$ & 3 &-1/1152 & -1/576 \\
888 &-$\sqrt{3}/3$ & 1 &-1/288  & -1/144 \\
\end{tabular}
\end{ruledtabular}
\end{table}

One can calculate the matrix elements of  ${\mathcal C}_{ijk}$ either from Eq. (\ref{3Q})
or from Eq. (\ref{QQQ}). The matrix elements of  ${\mathcal C}_{ij\overline k}$  can be
calculated either from Eq. (\ref{2QQBAR}) or from Eq. (\ref{QQQBAR}).

The first method is more tedious but it provides a check for
the second. It implies to apply successively the generators $F^{a}_{i}$ or  $\overline{F}^{a}_{}$
on the states  $|1 \rangle$, $|2 \rangle$ or $|3 \rangle$.
As an example, in Table \ref{detail} we show the contribution of all 
nonvanishing terms to Eqs. (\ref{3Q}) and (\ref{2QQBAR}). Their sums 
gives  the expectation value of ${\mathcal C}_{124}$ and of  ${\mathcal C}_{12\overline 5}$ 
respectively, for the color wave function $|3 \rangle $ defined in Eq. (\ref{ASYMPT}).
By taking into account the multiplicity  $m^{abc}$ of each term one  obtains - 5/36 
consistent with Eq. (\ref{QQQ}) and 5/18 consistent with Eq. (\ref{QQQBAR}), respectively. 
The latter value  is exhibited in column 2 of Table \ref{Table1}.

Thus, the third column gives the expectation value of the cubic invariant ${\mathcal C}_{124}$ 
for the state $|3 \rangle$. It is very easy to calculate  expectation value of ${\mathcal C}_{123}$ 
for the asymptotic state $|1 \rangle$ using its definition (\ref{SINGLET}). The result will coincide
with the column 3 and for symmetry reasons  all ${\mathcal C}_{ijk}$ are identical. As a matter of fact
the result is also identical to column 3 of Table II of Ref.  \cite{Pepin:2001is}
where a particular singlet color  state of six quark systems has been taken as an example.
This is a state which can be rewritten as a part of a $q^4 \bar q$ wave function where the 
subsystems $q^3$ and $q \bar q$ are octets in the color space. 

\vspace{1cm}

\centerline{\bf Acknowledgments}

The author acknowledges support from the Fonds de la Recherche Scientifique - FNRS under the
Grant No. 4.4501.05.


\end{document}